\documentclass[fleqn,letters,usenatbib]{mnras}

\usepackage{newtxtext,newtxmath}

\usepackage[T1]{fontenc}

\usepackage{graphicx}	
\usepackage{amsmath}	

\title[Pre-processing out to the turnaround radius of clusters]{The Role of Groups in Galaxy Evolution: compelling evidence of pre-processing out to the turnaround radius of clusters}

\author[Lopes et al.]{
Paulo A. A. Lopes,$^{1}$\thanks{E-mail: plopes@ov.ufrj.br}, Andr\'e L. B. Ribeiro$^{2}$, Douglas Brambila$^{1}$
\\
$^{1}$Observat\'orio do Valongo, Universidade Federal do Rio de Janeiro, Ladeira do Pedro Ant\^onio 43, Rio de Janeiro, RJ, 20080-090, Brazil\\
$^{2}$Laborat\'orio de Astrof\'isica Te\'orica e Observacional -- Departamento de Ci\^encias Exatas e Tecnol\'ogicas --Universidade Estadual de Santa Cruz,\\ 45650-000, Ilh\'eus, BA, Brazil\\
}

\date{Accepted XXX. Received YYY; in original form ZZZ}

\pubyear{2023}

\begin{document}
\label{firstpage}
\pagerange{\pageref{firstpage}--\pageref{lastpage}}
\maketitle

\begin{abstract}

We present clear and direct evidence of the pre-processing effect of group galaxies falling into clusters in the local Universe ($z \lesssim 0.1$). We start with a sample of 238 clusters, from which we select 153 with N$_{200} \ge$ 20. We considered 1641 groups within the turnaround radius ($\sim$ 5$\times$R$_{200}$) of these 153 clusters.
There are 6654 {\it individual cluster galaxies} and 4133 {\it group galaxies} within this radius. We considered two control samples of
galaxies, in isolated groups and in the field. The first comprises 2601 galaxies within 1606 {\it isolated groups}, and the latter has 4273 field objects. The fraction of star forming galaxies in infalling groups has a distinct clustercentric behavior in comparison to the remaining cluster galaxies. Even at $5 \times $R$_{200}$ the {\it group galaxies} already show a reduced fraction of star forming objects. At this radius, the results for the {\it individual cluster galaxies} is actually compatible to the field. That is strong evidence that the group environment is effective to quench the star formation prior to the cluster arrival. The group star forming fraction remains roughly constant inwards, decreasing significantly only within the cluster R$_{200}$ radius. 
We have also found that the pre-processing effect depends on the group mass (indicated by the number of members). The effect is larger for more massive groups. However, it is significant even for pairs an triplets. Finally, we find evidence that the time scale required for morphological transformation is larger than the one for quenching.

\end{abstract}




\section{Introduction}

According to the concordance model of the Universe ($\Lambda$CDM) low-mass dark matter halos are formed first (at high redshift), while larger halos come later through mergers and/or accretion of smaller systems. In this hierarchical structure formation scenario galaxy clusters represent the most massive and latest systems to form in the Universe due to their own gravity. Hence, we expect the presence of substructures, in the form of infalling groups, within clusters. That has been detected for many years, in different wavelengths \citep{bah77, jon84, dre88, moh93, pin96, gir97, lop06, lop18}.

Galaxy properties are well known to depend on the environment they are located \citep{oem74, dre80, bal06, cuc06, coo06}. Late-type, gas rich, blue star forming galaxies prefer sparse populated regions of the Universe, while early-type, gas poor, red passive objects dominate the most dense locations, such as the central parts of groups and clusters. Several mechanisms are expected to influence galaxy evolution in dense environments. Those processes can be related to interactions to other members and/or the cluster potential. Another possibility is through interactions with the hot gas trapped in massive systems (groups and clusters, \citealt{del12}). However, some mechanisms (such as tidal and ram pressure stripping) are more effective in the central parts of clusters, while some are more common in their outskirts and within groups. For example, due to the high relative velocity between cluster members, mergers are a rare phenomenon within their virial radius, being more common inside groups.

The combination of the hierarchical structure growth with the environment dependence of galaxy properties naturally creates the expectation that part of galaxies within clusters (those infalling within groups) are more evolved than the remaining. This fast aging is a result of their prior life within the dense environment of a group. The term used to describe this phenomenon is 'pre-processing' \citep{zab96, fuj04}. This process has been extensively investigated in the past years, using simulations \citep{bah19, han18, bak21} and/or observations \citep{cor06, dre13, rob17, ein20, est23}. It can be detected through the investigation of many different properties. For instance, it has been shown that the fraction of star forming (SF) galaxies in clusters increases from the center to the outskirts, but never reaching the field level (even at very large radius). The interpretation is that galaxies arriving in clusters, but previously belonging to groups, had their evolution accelerated in the group environment, leading to a reduced fraction of SF objects among those systems \citep{lew02, hai15, bia18}. The existence of merging features among some cluster galaxies is also taken as an indirect proof of the pre-processing effect, as mergers are much more likely to happen within groups. It is important to bear in mind that galaxies arriving in clusters within groups will also face what is called {\it post-processing}, a combination of the environmental effects from the parent group and the cluster. However, it is difficult to disentangle this effect, which has not been deeply investigated (an exception is found in \citealt{cho19}).

Previous studies in the literature have focused on different aspects of the pre-processing effect. For instance, \citet{mcg09, del12, bah13, pal19} investigated the accretion history of group galaxies into clusters. \citet{don21} aimed to disentangle the effects of AGN feedback, environment, and pre-processing, finding those depend on the galaxy and host mass. Some works investigated the variation of the star-forming (or passive) population as a function of clustercentric distance \citep{hai15}. However, just a few studies tried to separate the group from the non-group populations \citep{hou14, bia18} and sampled clusters down to at least 5$\times$R$_{200}$ \citep{lew02}.

This letter presents an investigation of the pre-processing effect (focusing on the variation of the star formation and late-type fractions), based on a large sample of cluster galaxies, which are separated in galaxies belonging or not to infalling groups, up to the turnaround radius. We have also used two control samples of galaxies, in isolated groups and in the field, for comparison. 

This work is structured as follows. In Section 2 we described our data and methodology to build each sample. In $\S$3 we present our results, while a discussion is made in $\S$4. The cosmology assumed in this work considers $\Omega_{\rm m}=$0.3, $\Omega_{\lambda}=$0.7, and H$_0 = 100$ $\rm h$ $\rm km$ $s^{-1}$ Mpc$^{-1}$, with $\rm h$ set to 0.7.

\section{Data and Methodology}

\subsection{The galaxy data}
\label{gals}

The photometric and spectroscopic data used in this paper were taken from the seventh release of the Sloan Digital Sky Survey (SDSS). The magnitudes retrieved from the SDSS are de-reddened model magnitudes. We derived absolute magnitudes taking in account the distance modulus, k and e$-$corrections. Rest-frame colours are also derived for all objects. We also use the total stellar mass and star formation rate values (SFRs) obtained by the MPA-JHU group \citet{bri04}. Note that this study is based only on a complete sample of what we call {\it bright galaxies}, having M$_r \le \text{M}^* + 1$ ($\le$ -20.58). We also impose a minimum stellar mass cut (Log M$_*$ = 9.50).

We adopt the $\Sigma_5$ galaxy density estimator as a tracer of the local environment. 
For each galaxy in our sample, we compute the projected distance, d$_5$, to the 5th nearest galaxy around it. We also impose to the neighbor search a maximum velocity offset of 1000 $km$ $s^{-1}$, and a maximum luminosity, which we adopt as M$_r = \text{M}^* + 1$. The local density $\Sigma_5$ is simply given by 5/$\pi$d$_5^{2}$, and is measured in units of galaxies/Mpc$^2$. Finally, we also take in account the fiber collision issue when deriving galaxy densities. The procedure is well described in \citet{lab10, lop14}.
  
\subsection{Clusters and groups}
\label{cls_grps}

This work investigates the properties of galaxies belonging to clusters out to the turnaround radius (R$_{\text{ta}}$, assumed to be $\sim$ 5$\times$R$_{200}$, as shown in \citealt{rin06}). Some of these galaxies are falling into the clusters as part of other systems (groups), while the remaining comprises the rest of the cluster population (galaxies not associated to any infalling group). For the current work, we call the first population as {\it group galaxies} and the second as {\it individual cluster galaxies}. 
We separate the two populations as we identify which cluster galaxies belong to infalling groups. We describe below the cluster and group samples and the identification of the different galaxy populations.

\subsubsection{The cluster sample}
\label{clusters}

The cluster sample is a combination of different catalogs that we have been working with for the past years. We have clusters from the supplement version of the Northern Sky Optical Cluster Survey (NoSOCS, \citealt{lop04, lop09a}), the Cluster Infall Regions in the SDSS (CIRS, \citealt{rin06}), the HIghest X-ray FLUx Galaxy Cluster Sample (HIFLUGCS, \citealt{rei02, and17}), the Planck Early Sunyaev-Zel'Dovich (ESZ, \citealt{pla11}), the SPIDERS catalog \citep{kir21} and now we also add clusters from \citet{tem12} (see $\S$\ref{groups} below). 

In \citet{bra23} we combined all the catalogs above with the exception of the one from \citet{tem12}. For the current work we consider only objects with 0.03 $ \le z \le $ 0.10. The upper redshift limit ($z = 0.10$) is due to the completeness limit of the SDSS main spectroscopic sample, limited at $r_{\text{petro}}$ = 17.77. That corresponds to an absolute magnitude limit of M$_r \sim \text{M}^* + 1 = -20.58$ at $z \sim 0.10$. \footnote{Note those are the cluster redshift limits. The galaxies within clusters obviously span a slightly broader range ($0.025 < z < 0.105$).} We consider as clusters all objects with M$_{200} \ge 10^{14}$ M$_\odot$. Given these constraints, we have 133 clusters from the above catalogs (without those from \citet{tem12}).


We applied the shifting gapper technique \citep{fad96, lop09a} to all galaxies with available redshifts around each cluster to select members and exclude interlopers. One difference to our previous approach is to consider all galaxies within $10.0$ h$^{-1}$ Mpc (instead of $2.5$ h$^{-1}$ Mpc), as we now want to have a member list sampling the infall pattern of the clusters. However, we only use the members within $2.5$ h$^{-1}$ Mpc to derive an initial estimate of the velocity dispersion ($\sigma_{\text{cl}}$). An estimate of M$_{200}$ is obtained adopting the equation 1 of \citet{fer20} (also see \citealt{mun13}). The corrections considered by \citet{fer20} to $\sigma_{\text{cl}}$ and M$_{200}$ are also employed. Next, an estimate of R$_{200}$ is derived from the above mass estimate. Then we derive final $\sigma_{\text{cl}}$ and mass estimates, now considering only members within R$_{200}$ (instead of $2.5$ h$^{-1}$ Mpc). We refer the reader to \citet{lop09a, lop14, lop18} and \citet{fer20} for more details on the estimates above.

This same procedure is applied to all systems in the catalog from \citet{tem12}, with at least three FoF members and 0.03 $ \le z \le $ 0.10 (17801 objects). We only use the coordinates, redshift, and velocity limits (derived from the FoF members) as input to our code. We
are able to obtain estimates of $\sigma_{\text{cl}}$, R$_{200}$ and M$_{200}$ for 2854 groups and clusters of \citet{tem12}. We then kept only the 189 clusters of this data set (M$_{200} \ge 10^{14}$ M$_\odot$).

Our original sample described above has 120 clusters (out of 133) within the main contiguous area of SDSS, which is used by \citet{tem12} (see their Fig. 1). Hence, we combined these 120 clusters from our original list to the 189 from \citet{tem12}, resulting in 238 objects. We still impose a final cut requiring a minimum number of 20 galaxies within R$_{200}$ to call an object as a cluster (N$_{200} \ge$ 20), leading to a final cluster sample of 153 systems.

There are 17839 {\it individual galaxies} associated to the 238 clusters described above. Considering only clusters with N$_{200} \ge$ 20 and galaxies within 5 $\times$ R$_{200}$ we are left with 12628 galaxies within the 153 clusters. We actually only work with {\it bright galaxies} (M$_r \le \text{M}^* + 1$), and impose that Log M$_* \ge$  9.50, so that the final sample comprises 6654 cluster galaxies. A very important remark is that these numbers do not represent the actual number of members obtained by the shifting gapper method. The reason is that we actually excluded from this cluster member list galaxies that are also members of infalling groups (4133 objects), as described in $\S$\ref{infall_groups}. Hence, the total number of cluster members is larger (10787), as we split those in two populations, of {\it group galaxies} and {\it individual cluster galaxies}. The number above (6654) reflects only the second population.

\subsubsection{The group sample}
\label{groups} 

We adopted the group sample from \citet{tem12}, which was built using Sloan Digital Sky Survey (SDSS) Data Release 8 (DR8). In reality, their sample comprises groups and clusters. Their clusters were actually use in combination to the other cluster samples above (see $\S$\ref{clusters}). However, the only group sample we consider for this work is theirs. 

The authors applied a modified friends-of-friends (FoF) method with a variable linking length in both directions, eliminating selection effects and achieving high completeness. Their catalog has 77858 groups (and clusters) with two or more members. The number of observed members in each group is called its richness. The authors also provide other group parameters that scale with mass, such as an estimate of the virial radius (given by the the projected harmonic mean), the velocity dispersion and total luminosity of the group in the $r-$band. Further details on the generation of their catalog with the FoF algorithm, and the avaliable galaxy and group properties, can be found in \citet{tem12}.

There are 40330 objects in the catalog from \citet{tem12}, with 0.03 $ \le z \le $ 0.10. In order to eliminate common clusters to our list with 238 clusters described above (in $\S$\ref{clusters}), we compared the two catalogs, keeping 40100 systems with at least two members from \citet{tem12}. Those objects have no counterpart to the 238 clusters. For these groups, we kept the FoF membership assignment, which is more appropriate to this very low number of members.


\paragraph{The infalling group sample}
\label{infall_groups}
\
\newline

We considered a membership matching approach to select groups in the infall regions of clusters. We compared all the galaxies that are group members (according to the FoF) to those that are cluster members (from the shifting gapper technique). All the FoF groups with galaxies matched to the cluster member list are considered as infalling groups. 

It is obviously possible that not all galaxies from an infalling group have a counterpart among the cluster members. That happens for groups that are close to the escape velocity of a cluster at a given radius. That could be the case for recent arrivals into the clusters or due to imperfect membership assignment (from the FoF or shifting gapper). For those cases, we did consider all group galaxies in our analysis, even if some of them are not matched to cluster galaxies. These objects correspond to $\sim 12 \%$ of the group galaxies. However, we verified that our results ($\S$\ref{pre_process}) are not affected if we exclude them.

It is important to keep in mind that what we consider as groups infalling into clusters are systems located within the caustic profile associated to the cluster members. This profile can be seen as the result of the membership selection produced with the shifting gapper technique. Although we have infalling groups with small clustercentric distances ($\lesssim$ R$_{200}$), their velocity offset distribution (relative to their parent clusters) is non-Gaussian, as expected from an infalling population. That is in full agreement with \citet{hai18}, who selected infalling X-ray groups around massive clusters. Due to the limited field of view their selection resulted only in groups with small radial offsets ($0.3 \lesssim \text{R/R}_{200} \lesssim 1.3$), but with a non-Gaussian velocity distribution.

We find 3792 groups in the infall regions of clusters. However, as we consider the infall region limited to the turnaround radius ($\sim$5 $\times$ R$_{200}$) we actually have 1941 groups within this limit, being 1641 groups associated to clusters with N$_{200} \ge$ 20.

There are 18340 galaxies in the 3792 infalling groups. Considering only groups that are in the infall of clusters with N$_{200} \ge$ 20, we have 15240 galaxies, from which 7477 are within 5 $\times$ R$_{200}$ of their parent cluster. Out of those we have 4133 bright galaxies (also having Log M$_* \ge$  9.50). Those are the members of our infalling group sample of 1641 systems.

\paragraph{The isolated group sample}
\label{isolated_groups}
\
\newline

We have also created a sample of isolated groups. We did so to have a comparison group sample free of the effects related to the cluster environment. We have 36308 groups that are not in the infall patterns of clusters (after excluding the 3792 infalling groups from the sample of 40100 groups). To consider a group as isolated we do the following. We compare each group to high density galaxies ($\Sigma_5 \ge$ 5 gals/Mpc$^2$; typical for galaxies within R$_{200}$ of clusters). A group is considered isolated if it is not found within 15 Mpc and $\pm$ 3000 km s$^{-1}$ from a high density galaxy. From the 36308 groups we selected 1606 isolated groups. There are 4097 galaxies within those systems, from which 2601 are bright galaxies with Log M$_* \ge$  9.50.

\subsection{The field sample}
\label{field}

One way to characterize the pre-processing effect is through the comparison of the fraction of star forming galaxies (or other populations or galaxy properties) in clusters and in the field. We built a field sample through the comparison of each galaxy (from SDSS DR7) to a group and cluster catalog, as in \citet{bra23}. However, here we discard galaxies with a distance smaller than 4 Mpc and
with $ |\Delta z| \le 0.10$ of any object from a combined cluster catalogue (based on the sample from \citealt{gal09} and all catalogues described above). Besides that, we also avoided galaxies within a radius of 2.0 Mpc and velocity difference of $\pm$ 2000 km s$^{-1}$ from a bright neighbor (M$_r \le \text{M}^* + 1$) in the spectroscopic main sample of the SDSS DR7. We found 4598 bright field galaxies in the same redshift range of cluster galaxies ($0.025 < z < 0.105$). The distribution of LOG $\Sigma_5$ shows a sharp cut at -0.5. However, there are still some galaxies ($< 0.5 \%$) with larger density values, which we exclude. We also remove one galaxy with LOG $\Sigma_5 =$ -4.0. Hence, the field list was then reduced to 4551 bright galaxies (also having Log M$_* \ge$  9.50). We performed a last cleaning, discarding 278 objects that were part of a group from \citet{tem12}, so that the final field sample comprises 4273 bright galaxies. However, the exclusion of these objects do not impact the field results shown below.

To sum up, unless otherwise stated, those are the numbers of bright and massive galaxies (M$_r \le \text{M}^* + 1$; Log M$_* \ge$  9.50) this work is based on: 6654 ({\it individual cluster galaxies}), 4133 {\it group galaxies}, 2601 {\it isolated group galaxies} and 4273 field objects. The median values of Log M$_*$ (the first and third quartiles are in parenthesis) for these four samples are 10.66 (10.48; 10.87), 10.65 (10.45; 10.87), 10.64 (10.43; 10.87) and 10.54 (10.33; 10.76), respectively.

\section{Pre-processing galaxies in groups}
\label{pre_process}

The pre-processing effect is usually characterized in the literature by the difference in the fraction of star forming galaxies (F$_{\text{SF}}$) in clusters and the field. We expect that the F$_{\text{SF}}$ in clusters to increase from the center to the outskirts of clusters. However, the measurements of F$_{\text{SF}}$ are generally found to be below the field results, even at large cluster radius ($\gtrsim $ 3 $\times $~ R$_{200}$, \citealt{hai15}). This difference is interpreted as the result of the group environment affecting cluster galaxies prior to their infall.

\begin{figure}
	\includegraphics[width=\columnwidth]{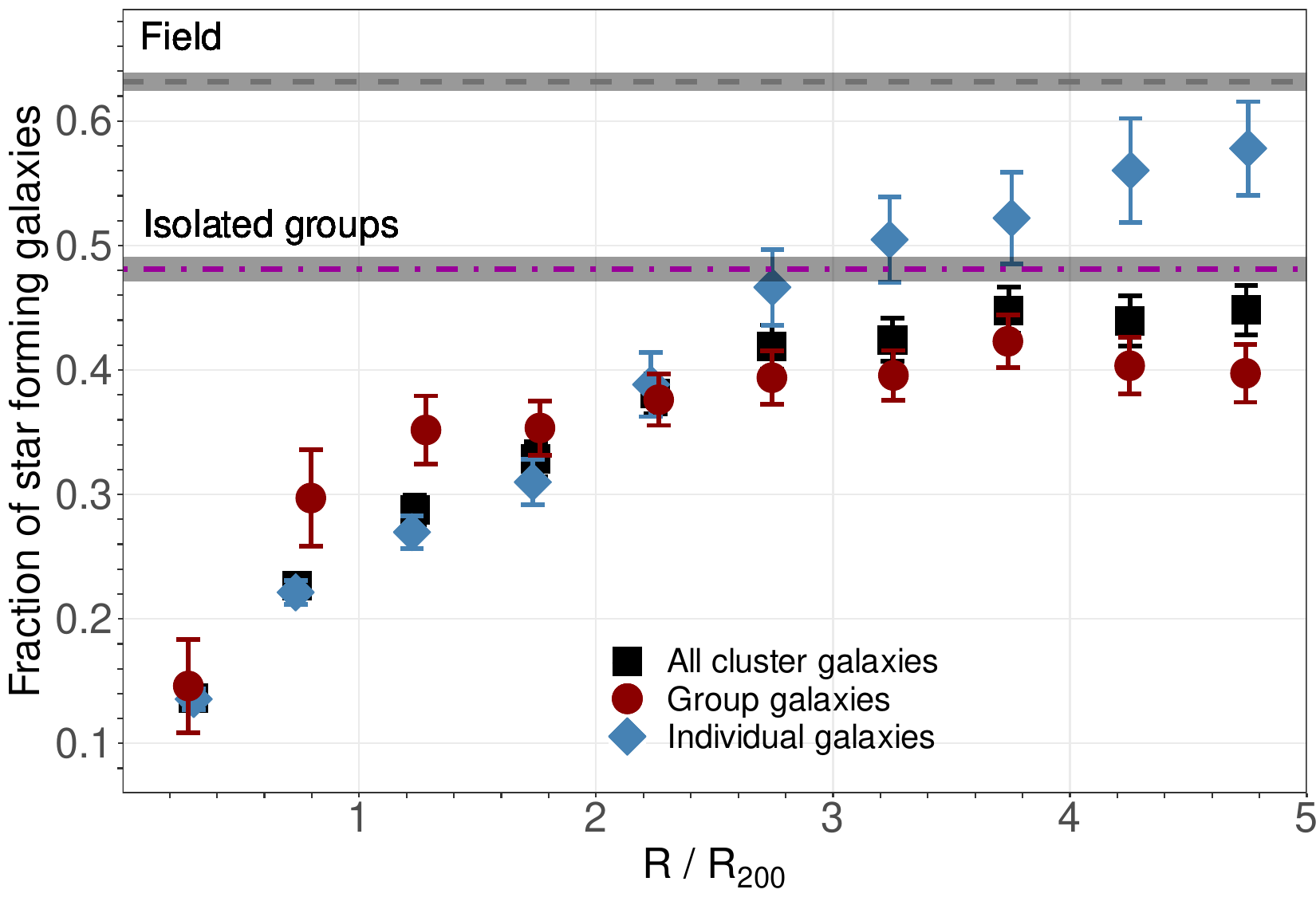}
    \caption{Fraction of star forming galaxies in clusters out to $5 \times$R$_{200}$. Galaxies that are not associated to groups are displayed in blue diamonds, while those that are members of infalling groups are shown in red circles. The fractions for all cluster galaxies (part of groups or not) are in black squares. The field fraction is shown by the gray dashed line, and the fraction for the isolated groups is displayed by the dot-dashed magenta line. F$_{\text{SF}}$ is computed in intervals of $0.5 ~\times$ (R $/$ R$_{200}$), with the values in the X coordinate given by the mean of all points within each interval. The error bars, and the gray bands over the two horizontal lines, indicate the 1$\sigma$ standard error of a proportion.}
    \label{fig:fsf_rad}
\end{figure}

In Fig.~\ref{fig:fsf_rad} we display the F$_{\text{SF}}$ as a function of clustercentric distance (up to 5 $\times $ ~R$_{200}$) for different populations. We consider the galaxy location in the star formation rate-stellar mass plane to classifiy galaxies as passive or star-forming. We call galaxies as SF if their log (SFR) value is greater or equal the line defined by equation 2 of \citet{tru20} (log SFR $= 0.70$ log M$_* - 8.02$). In Fig.~\ref{fig:fsf_rad} the black squares represent all galaxies that are cluster members and the gray dashed line indicates the field fraction. That is usually what we see in the literature (although observational results normally do not reach 5 $\times $ ~R$_{200}$), being the difference between field and clusters at large radii associated to the pre-processing effect. Here we further separate the cluster data in two subpopulations, of {\it individual cluster galaxies} (blue diamonds) and {\it group galaxies} (red circles); see $\S$\ref{clusters} and $\S$\ref{infall_groups}. We also show the results for galaxies in isolated groups (magenta dot-dashed line).

In addition to the usual indirect signature of the pre-processing effect (difference between the field and cluster results - gray line and black points) we now provide direct and compelling evidence in its favor. The F$_{\text{SF}}$ grows from the center to the clusters' outskirts. However, the blue diamonds ({\it individual cluster galaxies}) do not display a flat behavior for R $\gtrsim 2 \times $R$_{200}$. Actually, the F$_{\text{SF}}$ values of {\it individual cluster galaxies} are reconciled to the field fractions at $\sim 5 \times $R$_{200}$. The difference between the field and cluster results can be fully attributed to the {\it group galaxies} (red points), as those are previously pre-processed in these dense environments.

\begin{figure}
	\includegraphics[width=\columnwidth]{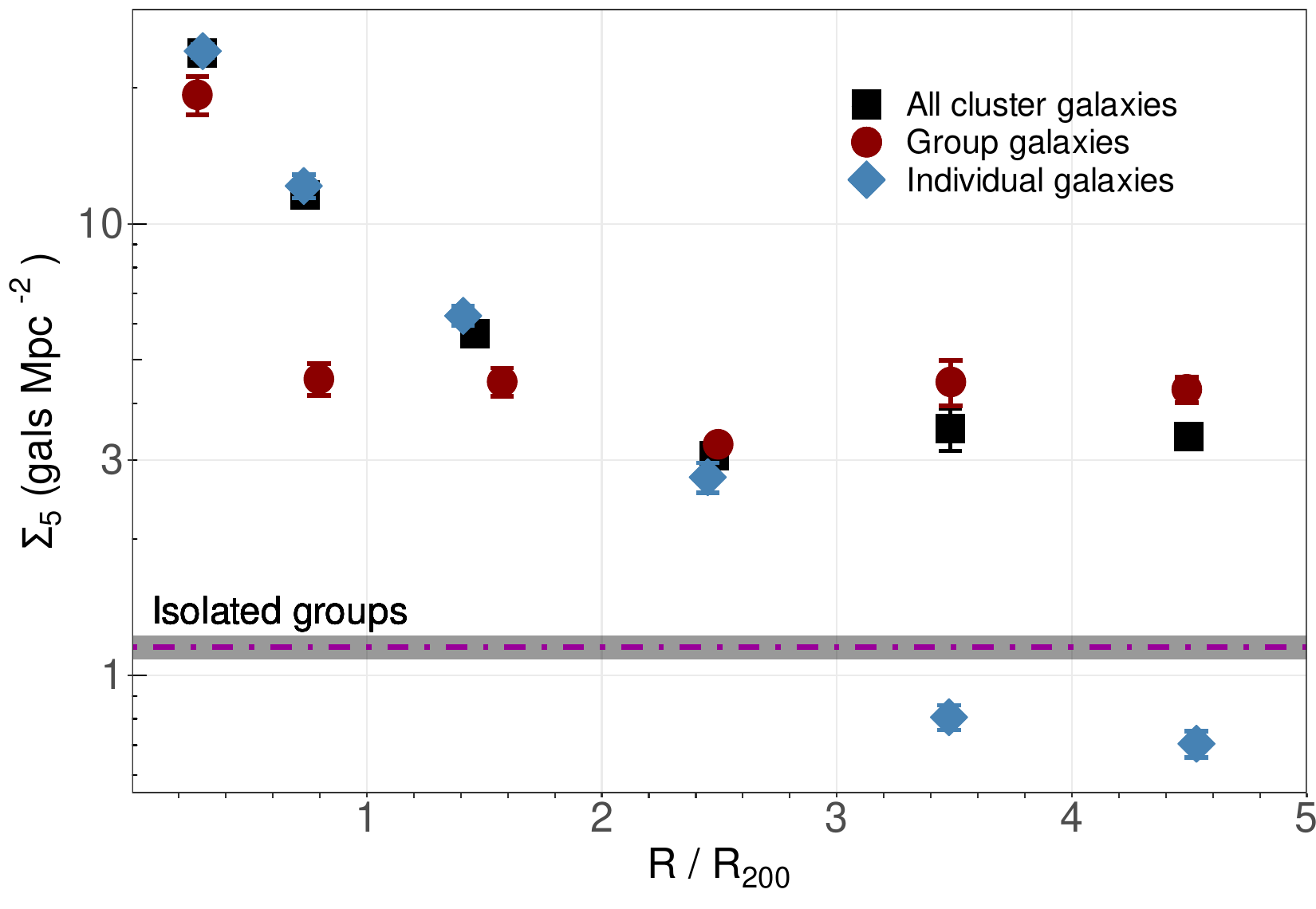}
    \caption{The correlation of local galaxy density with normalized clustercentric distance (R $/$ R$_{200}$). The colors of the symbols and lines are the same as for Fig.~\ref{fig:fsf_rad}. $\Sigma_5$ is computed in intervals of $1.0 ~\times$ (R $/$ R$_{200}$), with the values in the X coordinate given by the mean of all points within each interval. The exception is for the first two points (of each population) for which we consider intervals of $0.5 ~\times$ (R $/$ R$_{200}$). The error bars (and the gray band over the horizontal line) indicate the 1$\sigma$ standard error.}
    \label{fig:sig5_rad}
\end{figure}

Fig.~\ref{fig:sig5_rad} provides a natural explanation for the pre-processing effect. We display the local density parameter ($\Sigma_5$) {\it vs} (R $/$ R$_{200}$) for all galaxies in clusters (black squares), {\it individual cluster galaxies} (blue diamonds) and {\it group galaxies} (red circles). We
also show the results for the isolated groups (magenta dot-dashed line). The mean $\Sigma_5$ value for the field is 0.082 gals Mpc$^{-2}$. We omit it from the figure for clarity.
$\Sigma_5$ decreases from the central part of clusters to their outskirts. However, the {\it individual cluster galaxies} (blue points) reach much smaller densities (as they are not part of smaller systems), of $\sim 0.7$ gals Mpc$^2$. For R $\gtrsim 3 \times $R$_{200}$ the local density ($\Sigma_5$) becomes approximately flat, while the values for the {\it group galaxies} (red points) reach a plateau for R $\gtrsim 1-2 \times $R$_{200}$, with $\Sigma_5 \sim $ 4 gals Mpc$^2$, reflecting the higher density environment of groups. Hence, this plot, in combination with Fig.~\ref{fig:fsf_rad}, confirms that {\it group galaxies} are quenched before the rest of the cluster population ({\it individual cluster galaxies}), due to environmental effects within infalling groups.

\begin{figure}
	\includegraphics[width=\columnwidth]{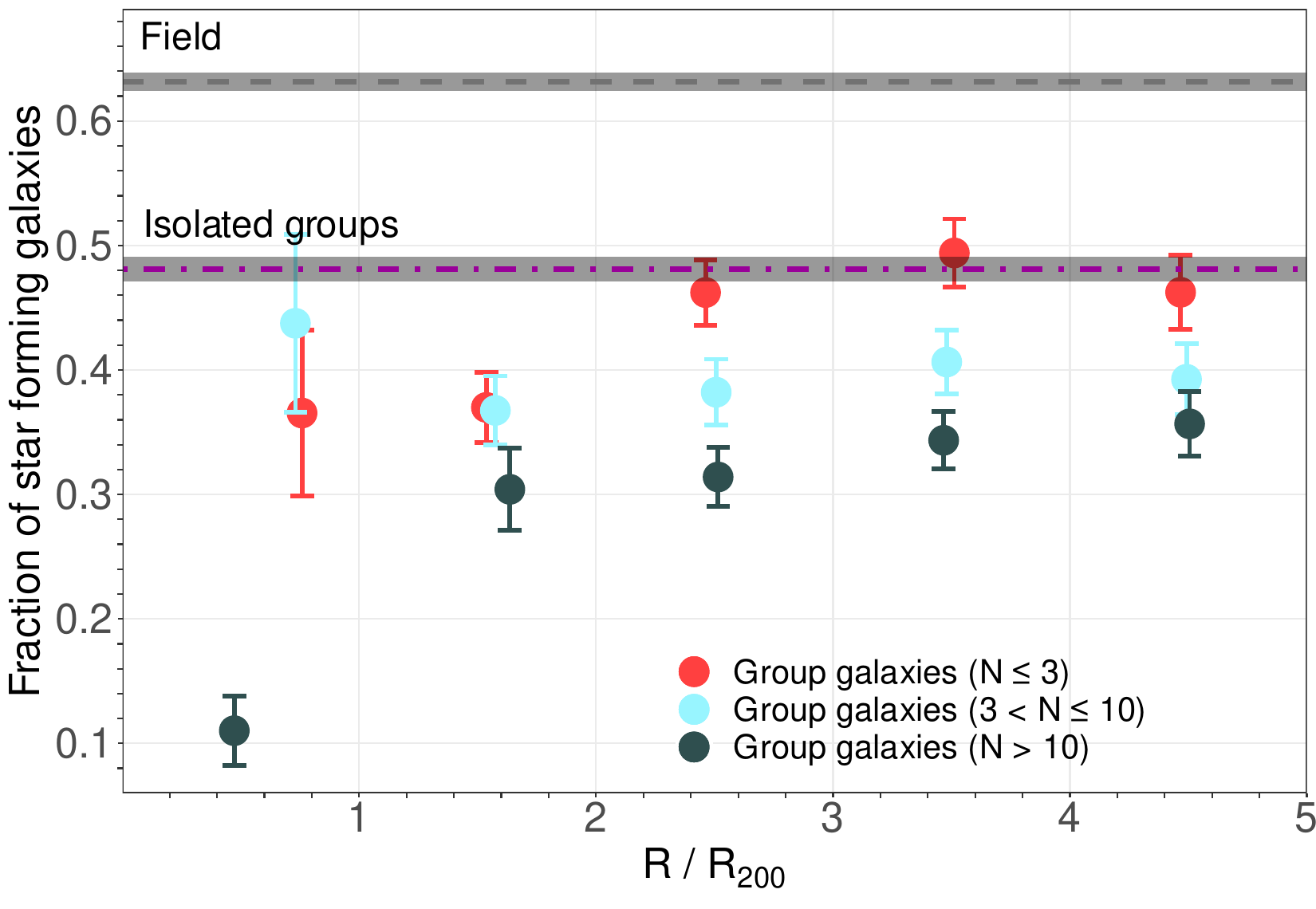}
    \caption{Fraction of star forming galaxies {\it vs} normalized clustercentric distance (R $/$ R$_{200}$) for galaxies within infalling groups. The results are divided according to the number of the members of the groups. Red points indicate the results for pairs and triplets (N $\le 3$), while the blue points represent systems with $3 < $ N $\le 10$, and dark gray points are for the most rich objects (with N $> 10$). The lines are the same as for Fig.~\ref{fig:fsf_rad}. F$_{\text{SF}}$ is computed in intervals of $1.0 ~\times$ (R $/$ R$_{200}$), with the values in the X coordinate given by the mean of all points within each interval. The error bars (and the gray bands over the two horizontal lines) indicate the 1$\sigma$ standard error of a proportion.}
    \label{fig:fsf_rad_ngroup}
\end{figure}

In Fig.~\ref{fig:fsf_rad_ngroup} we investigate the possible dependence of the pre-processing effect on group mass, traced by the number of members. The results for pair and triplets (N $\le 3$) are shown by the red points. The blue points display the fractions for objects with $3 < $ N $\le 10$, while we show in dark gray the results for groups with N $> 10$. The
results for the field (gray dashed line) and isolated groups (magenta dot-dashed line) are displayed as before. For the first subset (N $\le 3$) we detect a flat behavior from the outskirts down to $\sim 2 \times $R$_{200}$, with smaller fractions in the two inner bins. The second subset ($3 < $ N $\le 10$) displays a nearly flat behavior for the whole interval. The results for the most rich groups (N $> 10$) smoothly decrease towards the center (down to R $\sim 1 \times $R$_{200}$), but within R$_{200}$, F$_{\text{SF}}$ displays a steep drop. An important result is that at large radii (R $\gtrsim 2 \times $R$_{200}$) there is a significant difference in the F$_{\text{SF}}$ values according to the number of members. That indicates the fraction of quenched galaxies is already higher in large groups (in comparison to the smaller systems) when they arrive at the clusters. However, the pre-processing effect is detected even for pairs an triplets, as their F$_{\text{SF}}$ values are smaller than the field expectations.

\begin{figure}
        \includegraphics[width=\columnwidth]{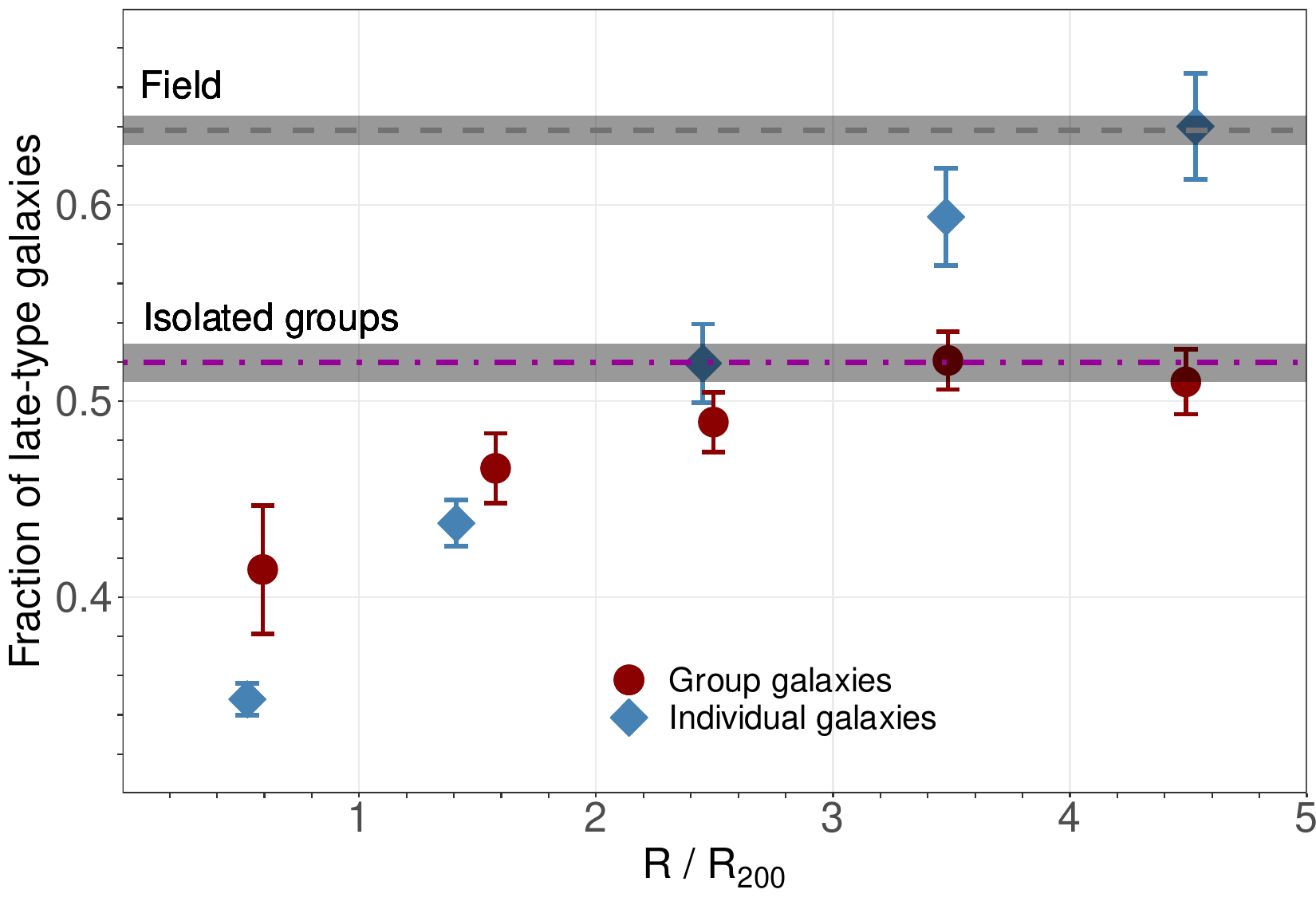}
    \caption{Analogous to Fig.~\ref{fig:fsf_rad}, but showing the fraction of late-type galaxies. The colors of the symbols and lines are the same as for Fig.~\ref{fig:fsf_rad}. F$_{\text{LT}}$ is computed in intervals of $1.0 ~\times$ (R $/$ R$_{200}$), with the values in the X coordinate given by the mean of all points within each interval. The error bars (and the gray bands over the two horizontal lines) indicate the 1$\sigma$ standard error of a proportion.}
    \label{fig:fLT_rad}
\end{figure}

The different clustercentric dependence, for the {\it group galaxies} and {\it individual cluster galaxies}, of F$_{\text{SF}}$ and $\Sigma_5$, is also seen for several structural and star formation related galaxy proprieties. For instance, similar results are found for galaxy specific star formation rate, color, size, concentration, T-type (a number assigned to each type of galaxy, which relates to the Hubble morphological sequence) and fraction of late-type (or early-type) galaxies. Those results reinforce the pre-processing effect and will be extensively discussed in a future paper. 
Here we show some results regarding morphology in Fig.~\ref{fig:fLT_rad}. To classify galaxies as early or late-type (LT) we consider the T-type values and probabilities derived by \citet{ds18}. We adopted the criteria described in \citet{bra23} (LT galaxies have T-type > 0, P$_{S0, bulge} < 0.6$ and P$_{disc} > 0.8$). We can see from Fig.~\ref{fig:fLT_rad} the fraction of late-type galaxies (F$_{\text{LT}}$) displays a similar behavior to F$_{\text{SF}}$ as function of (R $/$ R$_{200}$), corroborating the pre-processing in groups. The {\it group galaxies} have smaller values of F$_{\text{LT}}$ at large clustercentric distances, when compared to the {\it individual cluster galaxies}. However, an important result we achieve is the fact that the typical values of F$_{\text{LT}}$ are higher than F$_{\text{SF}}$, for both galaxy populations, suggesting that SF is quenched in a shorter time scale than what is necessary for the morphological transformation to happens (a result we previously showed in \citealt{lop14}). Note that our results are for bright massive galaxies. The F$_{\text{LT}}$ values at R $\sim 5 \times $R$_{200}$ of {\it individual galaxies} are compatible to their F$_{\text{SF}}$ results. But going inwards F$_{\text{SF}}$ decreases much faster than F$_{\text{LT}}$.

\section{Discussion and summary}

In this manuscript we present compelling and direct evidence in favor of the pre-processing effect of galaxies within groups. We do so through the comparison of the different populations of cluster galaxies, which we named {\it individual cluster galaxies} (not associated to any subgroup) and {\it group galaxies} (belonging to subgroups). This approach was tried only a few times in the literature (e.g., \citealt{hou14, bia18}). However, it is important to note that previous observational results usually do not reach the turnaround radius ($\sim 5 \times $R$_{200}$), being limited to at most 3 $\times $R$_{200}$ \citep{hou14, hai15, rob17, bia18}. An exception is the work of \citet{lew02}, but they do not split the cluster galaxies as described above. Some other results in the literature are also restricted to investigate the pre-processing effect in the surrounding regions of a single (or a few) cluster or within a supercluster region \citep{est23, ein20}. Our study is based on a large sample of groups (1641 infalling and 1606 isolated) and clusters (153), with the later sampled up to the R$_{\text{ta}}$. Within this radius we have 6654 {\it individual cluster galaxies} and 4133 {\it group galaxies}. Additionally, we have 2601 galaxies within {\it isolated groups} and 4273 field objects.

The F$_{\text{SF}}$ is measured separately for the two cluster populations all the way out to $5 \times $R$_{200}$ (see Fig.~\ref{fig:fsf_rad}; to the best of our knowledge that is shown here for the first time). A result first seen here is also the agreement between the cluster F$_{\text{SF}}$ infall ({\it individual cluster galaxies}) and field values, at large radius (close to the R$_{\text{ta}}$). That indicates that {\it individual galaxies} - when first arriving into clusters - do have similar fractions of star forming objects as in the field. As they travel inwards they are progressively quenched, even at large clustercentric distances (in agreement with \citealt{bah13}). On the contrary, the {\it group galaxies} arrive into the clusters with an already reduced fraction of star forming galaxies. Their F$_{\text{SF}}$ values are nearly constant inwards, down to $\sim 1-2 \times $R$_{200}$. A significant reduction is found only within R$_{200}$ (see the red points in Fig.~\ref{fig:fsf_rad}). 

An interesting result is that at $\sim$ R$_{200}$ the 
F$_{\text{SF}}$ of {\it group galaxies} becomes larger than the ones for the {\it individual cluster galaxies}. That is in disagreement with \citet{bia18}, who found comparable results within R$_{200}$. We interpret this inversion on the behavior of the two poulations (when going inside R$_{200}$) as a combination of two effects. First, the F$_{\text{SF}}$ results for the {\it individual cluster galaxies} (blue points) are {\it contaminated} by backsplash galaxies (that is expected in a large range around R$_{200}$), objects that already crossed the cluster cores once. Hence, being more affected by the cluster environment in comparison to the {\it group galaxies}. The latter are actually not expected to survive as group galaxies after the first cluster passage \citep{cho19, hag23}, so that the {\it group galaxies} shown here are probably recent arrivals into the clusters. However, this effect is expected to impact our results as well as the literature ones. Second, and most importantly, the results of Fig.~\ref{fig:fsf_rad} 
consider all groups in our sample. As seen in Fig.~\ref{fig:fsf_rad_ngroup} the F$_{\text{SF}}$ depends on the group mass (indicated by the number of members). Smaller groups (N $\le 10$) show much higher (especially within R$_{200}$) star forming fractions than the richer systems (N $> 10$). Note the results from \citet{bia18} should reflect what is expected for more massive systems, as they select groups through its X-ray emission.


We have also shown - perhaps for the first time - that the local galaxy densities ($\Sigma_5$, see Fig.~\ref{fig:sig5_rad}) of the two cluster populations have clustercentric variations that explain the F$_{\text{SF}}$ radial dependence. The $\Sigma_5$ values of the {\it individual cluster galaxies} decrease with radial growth, reaching much smaller values than obtained for the {\it group galaxies}, for R $\gtrsim 3 \times $R$_{200}$. The pre-processing effect is also verified through the reduced fractions of late-type galaxies in the {\it group} sample compared to the {\it individual cluster galaxies}. That is seen in Fig.~\ref{fig:fLT_rad}, from which we can also infer that the time scale required for the morphological transformation to be larger than the one for quenching.

We performed a few tests in order to verify the robustness of our conclusions. It is well known that the fraction of galaxy populations (such as star-forming or disc) and galaxy properties (such as color and morphology) are known to be correlated with the environment, but also stellar mass. The different populations investigated in the current work have different mass distributions. In order to check if that could impact our results we built stellar mass matched samples of the galaxies in different data sets. For instance, we verified that the F$_{\text{SF}}$ values of the {\it group galaxies} do not change significantly when using the stellar mass matched samples (in comparison to the {\it individual cluster galaxies}), which were built in radial bins. The variation of F$_{\text{SF}}$ looks less flat for R$ > 3 \times $R$_{200}$, but it is still compatible to the original one (seen in Fig.~\ref{fig:fsf_rad}). The only case the difference is larger than 1$-\sigma$ (but still within 2$-\sigma$) is for the last radial bin (close to $5 \times $R$_{200}$).

We have also investigated if the results change if we consider all galaxies in our original sample of 238 clusters, instead of the sample with 153 clusters (requiring N$_{200} \ge 20$). Another test we applied was to select only galaxies with Log M$_* \ge$ 10.0 (instead of 9.50). For both cases the F$_{\text{SF}}$ values become a little smaller, in comparison to our original results of Fig.~\ref{fig:fsf_rad}, but still in agreement, within the error bars.

The results presented in this study represent direct evidence that {\it group galaxies} are indeed quenched (and experience morphological transformation) before the rest of the cluster population ({\it individual cluster galaxies}), as the result of environmental processes within infalling groups. It is not the goal of the present work to investigate which processes are actually responsible for the accelerated quenching of the group galaxies before infall. Those could be related to galaxy encounters, starvation or ram-pressure stripping (in the case of the more massive groups), for instance. Our objective is to provide a careful selection of infalling groups within clusters and to disentangle the cluster population in two subsets: {\it group galaxies} and {\it individual cluster galaxies}. This way we are able to perform a detailed analysis of the variation of the F$_{\text{SF}}$ (and other properties) for these populations and to compare those to the results from the field and from an isolated group sample. Our results represent an important benchmark for cluster follow-up studies, out to $5 \times $R$_{200}$, aiming to investigate the pre-processing effect (e.g., the WEAVE Wide-Field Cluster Survey, \citealt{cor22}).

\section*{Acknowledgements}

P.A.A.L. thanks the support of {\it Conselho Nacional de Desenvolvimento Científico e Tecnológico} (CNPq), grants 433938/2018-8 and 312460/2021-0 and the {\it Fundação de Amparo à Pesquisa do Estado do Rio de Janeiro do Rio de Janeiro} (FAPERJ), grant E- 26/200.545/2023. ALBR thanks the support of CNPq, grant 316317/2021-7, and Fundação de Amparo à Pesquisa do Estado da Bahia (FAPESB) INFRA PIE 0013/2016. DB acknowledges the {\it Coordenação de Aperfeiçoamento de Pessoal de Nível Superior} (CAPES) for a PhD fellowship. This research has  made use of the SAO/NASA  Astrophysics Data System and the SDSS. A list  of participating institutions can  be obtained  from the  SDSS  Web Site
(http://www.sdss.org/).

\section*{Data Availability}

The galaxy, group and cluster catalogs used in this work are public available. However, we are willing to provide - upon reasonable request - the separate lists we have created.



\bibliographystyle{mnras}
\bibliography{preprocessing}








\bsp	
\label{lastpage}
\end{document}